\documentclass[a4paper,11pt]{article}

\usepackage{verbatim}
\usepackage{amsmath}
\usepackage{amsfonts}
\usepackage{amssymb}
\usepackage{mathtools}
\usepackage{physics}
\usepackage{graphicx}
\usepackage{tikz} 
\usetikzlibrary{decorations.pathreplacing,calc,calligraphy}
\usepackage{booktabs}
\usepackage{adjustbox}
\usepackage[utf8]{inputenc}
\usepackage{multirow}
\usepackage[splitrule]{footmisc}

\usepackage{pos}
\usepackage{empheq}
\usepackage{paralist}
\numberwithin{equation}{section}
\newcommand{\Lsp}{\ensuremath{\Lambda_\text{sp}}}
\newcommand{\Nsp}{\ensuremath{N_\text{sp}}}

\newcommand{\Mpld}{\ensuremath{M_{\text{Pl,}\, d}}}
\newcommand{\ellsp}{\ensuremath{\ell_{\text{sp}}}}

\newcommand{\ellpld}{\ensuremath{\ell_\text{Pl,d}}}

\newcommand{\ms}{\ensuremath{m_\text{str}}}
\newcommand{\mt}{\ensuremath{m_{\text{t}}}}
\newcommand{\dn}{\ensuremath{\mathrm{d}_n}}
\newcommand{\mn}{\ensuremath{m_n}}

\newcommand{\MBH}{\ensuremath{M_\text{BH}}}
\newcommand{\RBH}{\ensuremath{R_\text{BH}}}
\newcommand{\SBH}{\ensuremath{S_\text{BH}}}
\newcommand{\Sstr}{\ensuremath{S_\text{str}}}
\newcommand{\Lstr}{\ensuremath{L_\text{str}}}
\newcommand{\gsd}{\ensuremath{g_{\mathrm{str},d}}}

\title{A short overview on the Black Hole-Tower Correspondence and Species Thermodynamics}
\author*[a]{Alvaro Herr\'aez}
\author[a,b]{Dieter L\"ust}
\author*[a]{Joaquin Masias}
\author*[a]{Carmine Montella}
\renewcommand{\printHeadAuthors}{ }

\affiliation[a]{Max-Planck-Institut f\"ur Physik (Werner-Heisenberg-Institut),
Boltzmannstr. 8, 85748 Garching, Germany}
\vspace{0.1cm}
\affiliation[b]{Arnold Sommerfeld Center for Theoretical Physics, Ludwig-Maximilians-Universit\"at M\"unchen, 80333 M\"unchen, Germany }

\emailAdd{aherraez@mpp.mpg.de}
\emailAdd{luest@mppmu.mpg.de}
\emailAdd{jmasias@mpp.mpg.de}
\emailAdd{montella@mpp.mpg.de}

\abstract{The breakdown of gravitational effective field theories is intimately connected to the emergence of infinite towers of light states near infinite-distance limits in field space. In string theory, up to duality frame, such towers arise from Kaluza-Klein or weakly-coupled critical string oscillator modes.
Motivated by the Black Hole-String Correspondence, we review a broader mechanism whereby black holes undergo a transition into a tower of light states, governed by the Quantum Gravity cutoff---known as the Species Scale. 
Building on these developments, the Black Hole-Tower correspondence aims to provide a unified thermodynamic framework that describes black hole entropy in terms of the spectrum of the lightest degrees of freedom across various perturbative regimes of quantum gravity theories. In those regimes, thermodynamic consistency of such transition imposes stringent constraints on the spectrum, in agreement with string theory predictions. This defines the basis of the so-called Species Thermodynamics.

In this review, we emphasize these recent advances and synthesize their implications, offering an overview of how the outlined correspondence, the species scale and related thermodynamic principles enhance our understanding of black hole entropy within the effective field theory framework.}

 \renewcommand{\hookAfterAbstract}{%
\par\bigskip
\textsc{Report Number: MPP-2025-114}
}

\FullConference{Proceedings of the Corfu Summer Institute 2024 "School and Workshops on Elementary Particle Physics and Gravity" (CORFU2024)\
12 - 26 May, and 25 August - 27 September, 2024\\
Corfu, Greece\\}

\tableofcontents

\begin{document}
\maketitle
\pagebreak

%%%%%%%%%%%%%%%%%%%%%%%
%%%%%%%%%%%%%%%%%%%%%%%

\section{Introduction and motivation}
\label{s:Intro}
Einstein’s theory of General Relativity has proven remarkably successful in describing gravitational interactions at low energies, or equivalently, at large distance scales. However, it is well known that this theory is non-renormalizable at the quantum level and must therefore be treated as an effective field theory (EFT), valid only up to a certain ultraviolet cutoff \cite{Donoghue:1994dn,Donoghue:2022eay} There is strong evidence that the very framework of field theory breaks down once quantum gravitational effects become significant. In other words, beyond this cutoff, the UV completion of gravity is no longer a local EFT but instead a fully-fledged theory of Quantum Gravity (QG).
In general, not all effective field theories admit an ultraviolet completion within a consistent theory of quantum gravity. Conversely, quantum gravity frameworks impose stringent constraints on which EFTs can arise as low-energy limits. Understanding the generic features of EFTs that can consistently emerge from quantum gravity lies at the heart of the Swampland Program \cite{Vafa:2005ui}. These features are typically formulated as conjectures or hypotheses, which can be tested and, in some cases, rigorously derived within concrete setups such as various corners of string theory, through the AdS/CFT correspondence, or using black hole physics (see \cite{Brennan:2017rbf,Palti:2019pca,vanBeest:2021lhn,Grana:2021zvf,Agmon:2022thq,VanRiet:2023pnx} for reviews on the subject).

A fundamental feature of quantum-gravitational theories%, increasingly supported by evidence, 
is the existence of  infinite towers of states in the spectrum, marking a stark departure from the behavior of conventional quantum field theories. This expectation, well-motivated and extensively studied within string theory, is formalized in the so-called Swampland distance conjecture \cite{Ooguri:2006in}. It posits that in any $d$-dimensional effective field theory, an infinite tower of states becomes asymptotically massless---measured in $d$-dimensional Planck units---as one approaches infinite-distance limits in moduli space (or more generally, in scalar field space).
In the context of string theory, this idea can be further refined by the so-called Emergent String Conjecture \cite{Lee:2019wij}, which constrains the nature of these light towers by proposing that they must fall into one of two categories: either Kaluza-Klein towers or towers of oscillator modes from a weakly coupled critical string. In this sense, the appearance of towers of states becoming light signals the existence of a finite range of validity---in field space---for the corresponding effective field theory.

One might expect that integrating in these states as they become light would suffice to define a new EFT with an extended domain of validity. However, since these towers consist of infinitely many degrees of freedom, their presence suggests the breakdown of any standard EFT description above a certain energy scale. This leads to the notion of a fundamental maximum cutoff for gravitational EFTs, commonly referred to as the quantum gravity cutoff or species scale \cite{Dvali:2007hz, Dvali:2007wp, Dvali:2009ks}.
As in any field-theoretic description, this (purely gravitational) cutoff must manifest itself in the structure of higher-derivative corrections within the gravitational sector of the EFT expansion. In recent years, significant attention has been devoted to understanding the nature of this quantity \cite{ vandeHeisteeg:2022btw, Caron-Huot:2022ugt,vandeHeisteeg:2023dlw,Castellano:2023aum, Caron-Huot:2024lbf,Calderon-Infante:2025ldq}. In particular, it has been emphasized that any well-defined gravitational EFT in a perturbative regime must admit an expansion of the form:\footnote{To be precise, it has been argued in \cite{Calderon-Infante:2025ldq} that such an expansion generally contains also an additional term which yields the general structure
\begin{equation}
S \supset \dfrac{M_{\text{Pl},d}^{d-2}}{2} \int d^d x \sqrt{-g} \left[ R + \sum_{n>2} \dfrac{\mathcal{O}_n(\mathcal{R})}{\Lambda_{\text{sp}}(\phi)^{n-2}} \right]+ \int d^d x \sqrt{-g}\, \sum_{n>2}\frac{{\mathcal{O}}_n(\mathcal{R})}{\mt(\phi)^{n-d}}\,,
\end{equation}
where $\mt$ is the lightest mass gap of the mentioned infinite tower of states. We will neglect the second summand for the rest of this review since it turns out not to be relevant for the black hole configurations that we will explore, as explained in section \ref{s:BHTower}.}
\begin{equation}
S \supset \dfrac{M_{\text{Pl},d}^{d-2}}{2} \int d^d x \sqrt{-g} \left[ R + \sum_{n>2} \dfrac{\mathcal{O}_n(\mathcal{R})}{\Lambda_{\text{sp}}(\phi)^{n-2}} \right] 
\end{equation}
where $\Lambda_{\text{sp}}$ denotes the quantum gravity cutoff---which generically depends on other scalar fields, collectively denoted by  $\phi$. 

This quantum gravity cutoff was originally defined by the implicit relation \cite{Dvali:2007hz}
\begin{equation}
\Lambda_{\text{sp}} = \frac{M_{\text{Pl},d}}{N_{\text{sp}}^{\frac{1}{d-2}}},
\end{equation}
where $N_{\text{sp}}$ denotes the number of light species---i.e. those below the cutoff $\Lambda_{\text{sp}}$. Notably, this scale captures the key idea that a gravitational EFT with a large number of light degrees of freedom breaks down at an energy scale parametrically lower than the $d$-dimensional Planck mass $M_{\text{Pl},d}$, in contrast to the naive expectation that $M_{\text{Pl},d}$ sets the UV cutoff for gravitational EFTs. 

As we review in section~\ref{s:prelims}, when the infinite tower is dominated by Kaluza-Klein modes---associated to the decompactification of a $p$-cycle---or by string oscillator modes, the species scale $\Lambda_{\text{sp}}$ reduces to the higher-dimensional Planck mass $M_{\text{Pl},d+p}$ or to the string scale $\ms$, respectively, as expected.

An alternative and complementary derivation of the species scale arises from black hole physics. In this context, $\Lambda_{\text{sp}}$ is interpreted as the inverse curvature associated with the smallest semi-classical black hole that may be consistently described within a $d$-dimensional gravitational EFT \cite{Dvali:2009ks, Dvali:2010vm, Cribiori:2022nke}. 
This connection becomes especially clear in terms of black hole entropy. A Schwarzschild black hole of radius $\Lambda_{\text{sp}}^{-1}$ has entropy
\begin{equation}
 \SBH\simeq \left( \frac{R_{\text{BH}}}{4 \ell_{\text{Pl},d}} \right)^{d-2} \xrightarrow{{\RBH\to \Lsp^{-1}}} S_{\text{BH}} \simeq \left( \frac{\Lambda_{\text{sp}}}{M_{\text{Pl},d}} \right)^{2-d} \simeq N_{\text{sp}}.
\end{equation}
Interestingly, this entropy does not just follow the usual area law; in this case, it also turns out to be proportional to the total number of species. It is remarkable that this notion turns out to be deeply related to the previous definitions, as the same quantum gravitational corrections to the EFT that arise due to the presence of the infinite tower of states also also affect the geometry of black hole horizons---particularly in the small black hole regime \cite{Cribiori:2022nke, Calderon-Infante:2023uhz}. 

In the weakly coupled string regime---where oscillators modes from fundamental strings constitute the infinite tower of light states---there is strong evidence that a correspondence between black holes and towers of string oscillator modes can take place for black holes of the size of the string scale. This is known as the Black Hole-String correspondence \cite{Susskind:1993ws,Horowitz:1996nw}, and its detailed understanding has been the subject of several works in the past and recently \cite{Horowitz:1997jc,Chen:2021dsw,Susskind:2021nqs,Brustein:2021cza,Bedroya:2022twb,Balthazar:2022szl,Balthazar:2022hno,Halder:2023kza,Emparan:2024mbp,Ceplak:2024dxm,Chu:2024ggi,Bedroya:2024igb,Chu:2025fko}.

From the broader perspective reviewed above, the string scale arises as one specific realization of the quantum gravity cutoff---namely when the tower of light states consists of the oscillator modes of a weakly coupled string. More generally, one may ask whether transitions between black holes and towers of species can more generally occur at the species scale, or equivalently, at any generic perturbative regime within the EFT, with the Black Hole-String transition being a special case of this broader phenomenon.

Originally motivated by effective field theory studies of the thermodynamics of black holes and towers of weakly-coupled, light species, the concept of Species Thermodynamics was proposed based on a correspondence between minimal black holes and towers of species in~\cite{Cribiori:2023ffn}, and subsequently explored and developed in~\cite{Basile:2023blg, Basile:2024dqq, Herraez:2024kux, aparici:2025xxx}.
It was later argued that the equivalence between black holes and parametrically infinite towers of species can strongly constrain both the mass spectrum and the degeneracies of the species particles. This result was found to be consistent with the Emergent String Conjecture. Namely, that only Kaluza-Klein towers or weakly-coupled critical strings oscillators can consistently undergo such a transition into a minimal black hole. Consequently, this provided a first bottom-up argument in support of the conjecture~\cite{Basile:2024dqq}. The similarity between the relation of black holes to towers of species and the well-established Black Hole–String correspondence was then noticed and developed in~\cite{Herraez:2024kux}. This led to an explicit formulation of the Black Hole–Tower correspondence, conceived in close analogy to the Black Hole–String case. This identification was based on a detailed thermodynamic analysis of specific towers of species whose behavior was shown to reproduce black hole thermodynamics in the appropriate regime---hence deriving the constitutive relations of species thermodynamics form standard thermodynamics. In doing so, the proposal also provided a formal framework supporting the bottom-up argument for the Emergent String Conjecture.

This review consists of the recent developments concerning  Species Thermodynamics and the correspondence between black holes and towers of species \cite{Cribiori:2023ffn, Basile:2023blg, Basile:2024dqq, Herraez:2024kux}, and is divided as follows. In section \ref{s:prelims} we revisit some preliminaries about species and black holes thermodynamics. In section \ref{s:BHTower} we discuss the Black Hole-Tower Correspondence and its relation to the Emergent String conjecture, presenting a bottom-up argument for the latter. Finally, in section \ref{s:speciest} we review the laws of Species Thermodynamics as motivated by the physics presented in the previous sections.

\section{Towers of Species and Black Hole Thermodynamics}
\label{s:prelims}
In this section, we review the key ingredients to define the Black Hole-Tower correspondence in section \ref{s:BHTower} and Species Thermodynamics in section \ref{s:speciest}, summarizing some of the main argument presented in \cite{Cribiori:2023ffn, Basile:2023blg, Basile:2024dqq, Herraez:2024kux}. 
We review the concept of species scale in the presence of towers of light states, and summarize key thermodynamic properties of towers of species and those of black holes.

\subsection{Towers of states and the species scale}
\label{ss:speciesscaleintro}
In the presence of a high number of light species, the species scale \cite{Dvali:2007hz, Dvali:2007wp, Dvali:2009ks} gives an upper-bound for the cut-off of gravitational EFTs. It takes the form \cite{Dvali:2007hz}
\begin{equation}
\label{eq:species_scale}
\Lsp = \frac{M_{\text{Pl},d}}{\Nsp^{\frac{1}{d-2}}},
\end{equation}
where $M_{\text{Pl,d}}$ is the $d$-dimensional Planck mass and $\Nsp\gg1$ counts the number of light species, in a way that we will specify below.  

Whenever these light species consist of towers of states with increasing mass, as those appearing generically along weak coupling limits in gravitational theories \cite{Ooguri:2006in,Lee:2019wij}, it turns out to be useful to use a quantum number $n$ to parametrize the mass spectrum, $\mn$, and degeneracy, $\dn$, of the tower.\footnote{Note that for any given relation between physical quantities such as mass and degeneracy, there exist infinitely many reparameterizations. In the absence of UV information specifying the microscopic interpretation of the quantum number $n$, the only meaningful data we can rely on is the relation between $\mn$ and $\dn$.} Hence, we may consider a generic tower of species characterized by a mass spectrum of the form
\begin{equation}
\label{eq:generalm_n}
\mn =  \mathrm{f}(n)\,  \mt \, ,
\end{equation}
where $\mt$ denotes the characteristic mass scale of the tower---which generically coincides with the mass gap. 
We restrict  to Kaluza-Klein and string oscillator modes in this section, since they have been argued to be the only possible kinds of towers (up to dualities) that arise in weakly coupled limits in gravitational EFTs \cite{Lee:2019wij}. We consider more general towers in section \ref{s:esc}, where a general bottom-up argument for the Emergent String Conjecture is provided. 

In the case of Kaluza-Klein towers associated to decompactification of $p$ isotropic extra dimensions with $p$-dimensional volume $\mathcal{V}_p\, M_{\mathrm{Pl,}d+p}^{ -p}$, it is possible to show using Weyl's law (see e.g., \cite{strauss2007partial}) that the spectrum can be effectively parameterized by\footnote{For the explicit case of a decompactification limit of $p$ toroidal internal dimensions, with sizes $m_{t,i}^{-1}$, this can be recovered from the familiar formula
\begin{equation}
    m(n_1,n_2,\dots,n_p) = \sqrt{n_1^2\,m_{t,1}^2+n_2^2\,m_{t,2}^2+\dots +n_p^2\,m_{t,p}^2}.
\end{equation} }
\begin{equation}
    \mn =n\,  \mt \,,\quad \dn = \mathrm{d}_0\,n^{p-1},
    \label{eq:kktower}
\end{equation}
where $\mt\simeq \mathcal{V}_p^{-1/p}$ and $\mathrm{d}_0$ can account for the presence of multiple towers.\footnote{Without loss of generality we can effectively parametrize by a single one with $\mn=n^{1/p}\, \mt$ and constant degeneracy \cite{Castellano:2021mmx}} 

The species scale associated to such tower can then be identified with the mass of the heaviest weakly-coupled state in the tower $m(N)$, labeled by its level, $N$, whose mass coincides with the species scale itself\footnote{Note that the definition \eqref{eq:species_scale},  is physically meaningful only up to $\mathcal{O}(1)$ factors that cannot be accounted for by the naive counting in the EFT. Thus, we will not keep track of these order one factors in the definition of the species scale, which are nevertheless subleading in the limit $\Nsp\gg 1$, for the remainder of the argument.} 
\begin{equation}
\label{eq:LspKKtower1}
    \mn=N \, \mt =\Lsp\, , \qquad\Nsp=\sum_{n=0}^N n^{p-1}\simeq \frac{p}{p+1} N^p + \dots
\end{equation}
where we have summed over the degeneracies $\dn$ of the tower of states up to level $N$ to obtain the total number of species. Using \eqref{eq:LspKKtower1} we can express the number of species and the species scale exclusively in terms of $\mt$, the Planck scale and the number of dimensions that decompactify in the limit, giving 
\begin{equation}
\label{eq:LspNsppolynomialtower}
 \Lsp = \mt^{\frac{p}{d+p-2}}M_{\text{Pl},d}^{\frac{d-2}{d+p-2}}\, , \qquad \Nsp=\left(\dfrac{\Mpld}{\Lsp}\right)^{d-2}\simeq \left(\dfrac{\Mpld}{\mt}\right)^{\frac{(d-2)p}{d+p-2}}\, .
\end{equation}
It can be seen by using the relation  $\Mpld^{d-2}=M_{\mathrm{Pl},d+p}^{d-2} \mathcal{V}_{p}$ that the species scale displayed in \eqref{eq:LspNsppolynomialtower} gives precisely the higher dimensional Planck scale, $M_{\text{Pl},d+p}$, for finite $p$.

It is tempting to consider the limit $p\to\infty$ in the above equations. It provides a finite result which reproduces a tower with $\Nsp=(\Mpld / \mt)^{d-2}$ states whose mass scale is at the cut-off, $\Lsp = \mt\,$ \cite{Castellano:2021mmx}. Notice that the dependence on the Plank mass is gone, and the scale of the tower is itself the species scale.
It has been argued that this limit represents an effective tensionless string limit $\mt = \ms$, where the species scales matches the Hagerdorn temperature and is consistent with the analysis from the scales suppressing higher-curvature corrections in gravitational EFTs \cite{vandeHeisteeg:2022btw, Caron-Huot:2022ugt,vandeHeisteeg:2023dlw,Castellano:2023aum, Caron-Huot:2024lbf, Calderon-Infante:2025ldq}. 

A more precise parametrizatoin of the tower of oscillator modes from a weakly coupled critical string is
\begin{equation}
\label{eq:stringtowerspectrum}
    \mn=\sqrt{n}\,  \mt\, , \qquad \dn= g(n) \ e^{c\sqrt{n}}\, ,
\end{equation}are
where the mass scale of the tower is $\mt=\ms$ and $g(n)$ is a monomial in $n$ that depends on the particular string theory under consideration but is irrelevant for the results. For such an exponentially degenerate tower, we perform the detailed analysis of the counting and the species scale in the thermodynamic limit in the following subsection.

\subsection{Thermodynamics of towers of species}
\label{s:thermo}
We now review the thermodynamics of towers of  species both in the microcanonical and in the canonical ensembles, emphasizing the \emph{frozen species limit} and related limits necessary to define the correspondence presented in section \ref{s:BHTower}.

\subsection*{Microcanonical Ensemble}
 As a thermodynamic ensemble,  it is possible to define an entropy and energy for generic weakly-coupled towers of light states.
 It is important to remind that we are considering a perturbative regime $\Nsp\gg1$, assuming that the amount of light states below the cut-off is parametrically large. Hence, the thermodynamics of such ensemble can be computed at fixed energy in the microcanonical ensemble by counting the number of tuples $\{k_\mathbf{s}\}$ of occupation numbers of each state and species.\footnote{For exponential degeneracies, energy fluctuations are not small in the thermodynamic limit, so the microcanonical counting cannot be applied. For this case we will refer to the canonical ensemble with partition function $\mathcal{Z}=    \sum_{n=1}^\infty e^{-\frac{\mn}{\Lsp}}\dn$ in the next subsection (see \cite{Herraez:2024kux} for more details).} 
 
 Following \cite{Basile:2023blg}, the species entropy $S_\text{sp} = \log D(E_\text{sp})$ is then the microcanonical entropy of this ensemble, where $D(E_\text{sp})$ counts the number of combinations of species with total mass $E_\text{sp}=\mt \,M$, where $M \gg 1$ is an integer. In order to compute it, we consider the auxiliary partition function
\begin{equation}
    Z(q) = \sum_M D(M) \, q^M = \prod_{n\leq N} \frac{1}{\left(1 - q^{\frac{\mn}{\mt}}\right)^{\dn}} \, ,
\end{equation}
which counts how many microstates have total energy $m \, M$, where $\mathcal{M} \gg 1$ is an integer. The entropy can then be read off expanding around the saddle points of the total degeneracy $D(M)$
\begin{equation}
    D(M) = \frac{1}{2\pi i} \oint \frac{dq}{q^{M+1}} \, Z(q) \, \longrightarrow \,\, S_{\mathrm{sp}}(M) \overset{M \gg \Nsp}{\sim} \Nsp + \mathcal{O}(\log(\Nsp))\,.
\end{equation}
Additionally, as a first approximation for the energy we can consider the minimal energy of the ensemble $E_{\mathrm{sp}}$ (i.e. neglecting any potential contribution from configurations with non-vanishing momentum), namely the sum of all the masses in the tower 
\begin{equation}
    E_{\text{sp}} = \sum_{n=1}^{N} \mn\, \dn = \sum_{n=1}^{N} \mt \, \mathrm{f}(n)\, \dn
\end{equation}
For the case of a Kaluza-Klein-like tower—cf. \eqref{eq:kktower}, the energy and entropy of such system are given by 
\begin{equation}
\label{{eq:microentropy}}
S \simeq \frac{p}{p+1}, \Nsp \,, \quad E_{\text{sp}} = \frac{p}{p+1} \, \Lsp^{3-d}\, + \text{corr}.
\end{equation}
This straightforwardly implies $E\simeq S \Lsp$. Interestingly, these quantities remain finite in the limit $p \to \infty$, effectively capturing some properties of towers of string oscillators. We will return to this point in section~\ref{s:BHTower}.

\subsection*{Canonical Ensemble}
We consider here the canonical description, consisting of a system of free particles, including a tower of light modes of the general form described around eq. \eqref{eq:generalm_n}, in a box of size $L$ at temperature $T$,\footnote{This is needed in order to have a well defined canonical ensemble in Minkowski space in the presence of gravity \cite{Atick:1988si}.} such that modes with masses $\mn\lesssim T$ can be excited. It is known that the entropy end energy of such thermodynamic system is given by
\begin{equation}
\label{eq:standardsethermo}
    S= N_T L^{d-1} T^{d-1}\,,\,
    \qquad E= N_T L^{d-1} T^{d},
\end{equation}
were $N_T$ is the  number of \emph{active} species from the tower at a given temperature, $T$. Such quantity turns out to have an intuitive expression for towers of polynomial degeneracy, $N_T=(T/\mt)^p$, but has a more subtle---but still well defined expression in the canonical ensemble--- for exponentially degenerate towers below the Hagedorn temperature (see \cite{Herraez:2024kux}, section 3.2). In particular, studying the \textit{frozen momentum limit} $\,T=L^{-1}$, turns out to be of particular interest \cite{Castellano:2021mmx, Herraez:2024kux}. In practice, this means that the momentum of the species---in the $d$ non-compact directions--- effectively freezes, and do not contribute to the entropy. It is also in this limit that the canonical and microcanonical descriptions presented here match. For the Kaluza-Klein and the weakly coupled string oscillator towers, eqs. \eqref{eq:kktower} and \eqref{eq:stringtowerspectrum} recover the entropy and total energy of the effective towers
\begin{equation}
\label{eq:standardsethermofrozen}
    S\simeq  N_T\, , \,\quad  E=T\, N_T\, .
\end{equation}
Furthermore, in the species limit $T\to\Lsp$, the number of active modes matches with the full number of weakly-coupled species in the tower $N_{T}\to \Nsp$ recovering
\begin{equation}
\label{eq:SandEspecies}
    S_{\text{sp}}\simeq \Nsp\,\, ,\, \qquad E_{\text{sp}}\simeq \Lsp\,\Nsp,
\end{equation}

\subsection{Thermodynamics of black holes and black branes}
\begin{figure}[hb] 
 \centering 
 \includegraphics[width=0.80\textwidth]{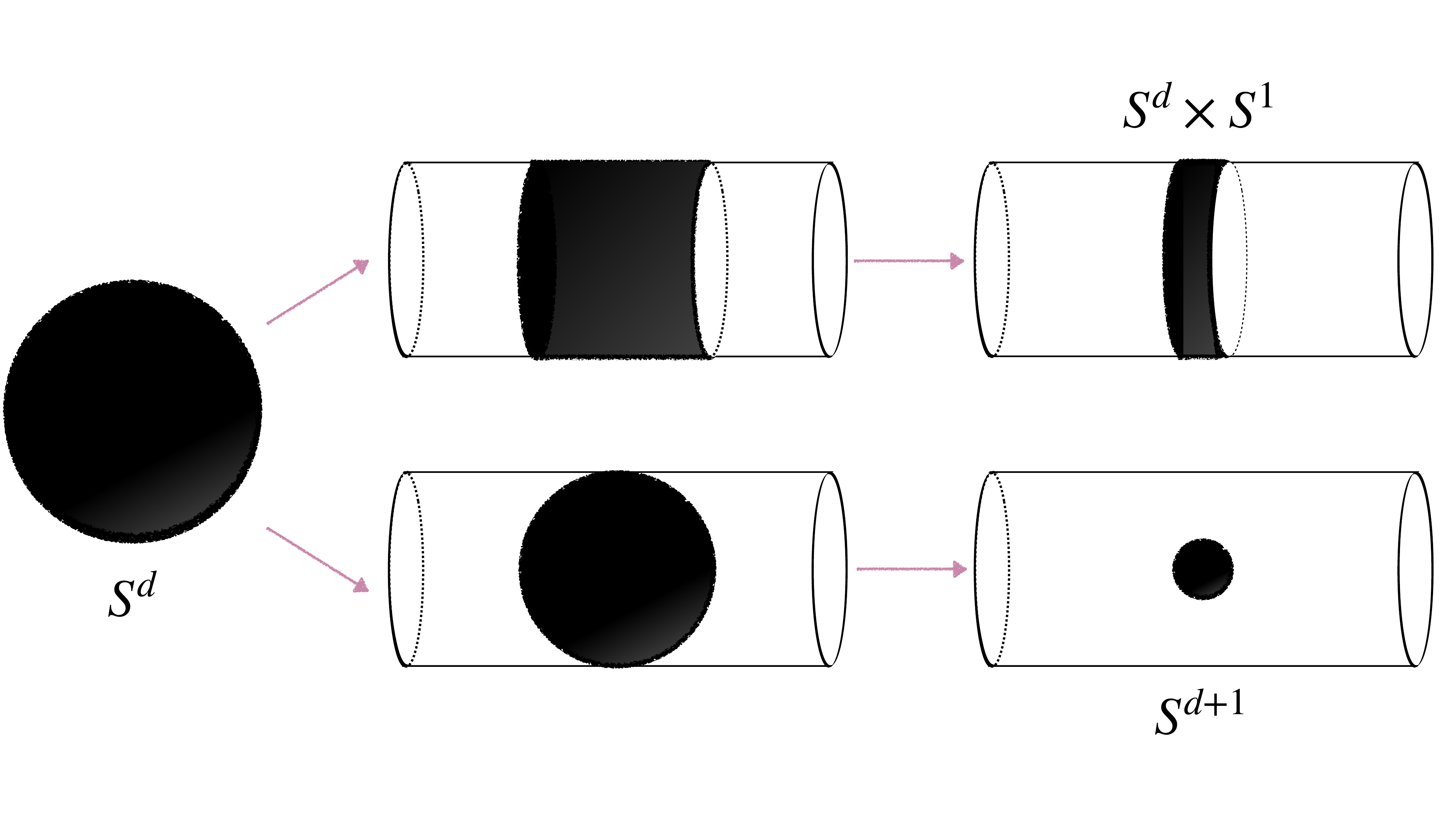} 
 \caption{Schematic representation of a $d$ dimensional black hole and its two possible $d+1$ counterparts as one increases the temperature, for a single compact dimension. The higher dimensional black hole is localized in the internal dimensions, while the black string wraps the internal $S^1$.} 
 \label{fig:BHBBscheme}
 \end{figure}

 Black holes are purely gravitational, non-perturbative objects, and their entropy occupies a central place in our understanding of quantum gravity. Starting with Bekenstein's insight that the entropy of a black hole scales with its horizon area and Hawking's seminal derivation of black hole radiation \cite{PhysRevD.7.2333,Hawking:1974rv}, one learns that gravitational systems encode information in a holographic manner \cite{tHooft:1993dmi,Susskind:1994vu,Maldacena:1997re}. More generally, the Covariant Entropy Bound (CEB) \cite{Bousso:1999cb,Bousso:1999xy} further suggested that the maximum entropy contained in a region is given by the area of the surface enclosing the region in Planck units\footnote{The covariant entropy bound formally states that the maximum entropy passing through light-sheets in a given spacetime region is proportional to the decrease in area of its boundary surface.}. These developments imply that, although black holes are classically described by only a few parameters, they account for an enormous number of microstates, a feature that we now explore in the presence of extra dimensions and weakly coupled stringy degrees of freedom.
In this section, we briefly highlight some key aspects of black holes and black branes in theories with $p$ extra dimensions with total volume $\mathcal{V}_p$. The characteristic length scale of the internal space is then $r=\mathcal{V}_p^{1/p}$. In particular, we discuss the correspondence and differences between what we denote as the \emph{minimal black hole in the EFT}—which for $\RBH<r$ we identify with a black brane solution wrapping the extra dimensions of size $r$—and the \emph{higher-dimensional black hole} solution, which for $\RBH<r$ is fully localized in the higher-dimensional spacetime, namely a $(d+p)$-dimensional spherical black hole localized both in the compact and non-compact directions.
This is depicted schematically in Fig. \ref{fig:BHBBscheme}

Considering neutral solutions with radius $\RBH$, two possibilities arise when the horizon size is smaller than the extra dimension scale $r$:

\begin{itemize}
    \item A higher-dimensional black hole localized in the full $D=(d+p)$-dimensional spacetime, with entropy scaling as
    \begin{equation}
    \label{eq:bhentropy}
    S_{\mathrm{BH},D}= \frac{4\pi^{D}}{\Gamma(D/2-1)} \RBH^{D/2} M_{\mathrm{Pl},D}^{D-2}\simeq  \MBH^{\frac{D-2}{D-3}}M_{\mathrm{Pl},d}^{-\frac{D-2}{D-3}} ,
    \end{equation}
    where $p$ is the number of internal dimensions.
    
    \item A black brane wrapping these $p$ extra dimensions, with entropy scaling as
\begin{equation}
\label{eq:bbentropytwo}
    S_{\mathrm{BB},p,D} =\frac{4\pi^{d}}{\Gamma(d/2-1)} R_{\text{BB}}^{d-2} \mathcal{V}_p M_{\mathrm{Pl},D}^{D-2}=M_{\mathrm{BB}}^{\frac{d-2}{d-3}}M_{\mathrm{Pl},d}^{-\frac{d-2}{d-3}} .
\end{equation}
\end{itemize}
Using the relation between Planck masses
\begin{equation}
M_{\mathrm{Pl},d}^{d-2} = \mathcal{V}_p M_{\mathrm{Pl},D}^{D-2},
\label{eq:planckmasses}
\end{equation}
these two solutions yield parametrically equivalent entropies when $\RBH = r$, making them indistinguishable from the point of view of an effective $d$-dimensional observer.

As first noted by Gregory and Laflamme \cite{Gregory:1993vy}, a non-extremal black brane presents a dynamical instability when $R_{\mathrm{BB}}<r$. Indeed, if the horizon is perturbative, the corresponding Lorentzian Lichnerowicz operator develops exponentially diverging modes, which makes the solution unstable. Heuristically, this can be seen from the fact that a localized black hole of the same mass, $\MBH = M_{\mathrm{BB}}$, is more entropic, which naively means that the higher-dimensional black hole is kinematically favored over the black brane.
 
One can however argue that the $d+p$ dimensional analogue of the $d$ dimensional black hole is indeed the black brane wrapping the internal directions. Using the relation between the Planck masses \eqref{eq:planckmasses}, one can see that the lower-dimensional black hole solution and the black $p$-brane one have parametrically the same entropy. We can compare EFT and black hole entropies in the specific limit of $\RBH\simeq 1/T\simeq 1/\Lsp\simeq 1/M_{\mathrm{Pl},D}$ and we see that the black brane (and the lower dimensional black hole) can be identified with the EFT that sees the presence of the extra dimensions. In contrast, a higher dimensional black hole that does not wrap the internal directions contains no information about the compact directions. This is to be expected, as the fully localized solution can only see the localized modes along the extra dimensions, as opposed to the wrapped black brane, which sees the internal geometry.\footnote{See also \cite{Bedroya:2024uva, Herraez:2025gcf} for a discussion of the Gregory–Laflamme instability in the context of the Swampland program.} It is important to distinguish that the process depicted in Fig. \ref{fig:BHBBscheme} is not a dynamical process, and in fact we do not consider the consequences of an unresolved Gregory-Laflamme instability (for a possible resolution see \cite{Emparan:2024mbp}), we limit ourselves to kinematical statements.

\section{The correspondence between black holes and towers of light states}
\label{s:BHTower}
Having discussed the thermodynamics of towers of (free) species and black holes, a natural question arises: can we use the information about the microscopic origin of the entropy of  towers of species to learn about the entropy of black holes? To answer that question, it important to keep in mind that in two important properties of gravitational theories. The first one is the existence of holographic entropy bounds \cite{Bousso:1999xy}, which in simple setups can be summarized in the idea that the maximum amount of information (or entropy) that can be stored in a closed spacelike region $\Sigma$, is bounded from above by the area enclosing the region $A(\partial \Sigma)$, measured in Planck units, instead of by its volume---as is the case in non-gravitational field theories. Equivalently, this can be stated as the fact that the maximal entropy that can be stored in a spherical region of size $L$ is bounded from above by that of the Schwarzschild black hole of the same size, whose entropy is precisely given by its horizon area in Planck units \cite{PhysRevD.7.2333} (c.f. eq. \eqref{eq:bhentropy})\footnote{In our conventions $G_{\mathrm{N,}d}=\ellpld^{d-2}=1/8\pi \Mpld^{d-2}$}
\begin{equation}
    S(\Sigma)\leq 2 \pi\  A(\partial\Sigma) \ \Mpld^{d-2} \simeq  L^{d-2} \ \Mpld^{d-2}\, .
\end{equation}

On the other hand, if we consider a thermodynamic system with total (average) energy $M$, and  temperature $T$, inside a spherical region of size $L$, it is known that the gravitational backreaction cannot be negligible if the total mass becomes too large \cite{Cohen:1998zx}. This can be estimated by the mass at which the corresponding Schwarzschild radius, $\RBH(M)$, would become of the order of $L$, since at that scale an outside observer would see the configuration as a black hole of said radius. This gives the bound
\begin{equation}
    L\gtrsim \RBH (M)=\left( \dfrac{M}{\Mpld}\right)^{\frac{1}{d-3}}\Mpld^{-1}\, .
\end{equation}
Using the standard thermodynamic relations \eqref{eq:standardsethermo} for (towers of) particles in a box of size $L$, we can translate both constraints into upper bounds for the energy and the entropy of the configuration, namely
\begin{align}
   &M\lesssim TL^{d-2} \, , \quad &S\lesssim L^{d-2} \, ,  \qquad &\mathrm{from} \mathrm{\ holographic \ entropy \ bounds} \label{eq:holographicbounds}  \\
&M\lesssim \,L^{d-3} , \quad &S\lesssim\dfrac{L^{d-3}}{T}\, , \qquad &\mathrm{from} \mathrm{\ avoiding\  gravitational \ collapse.}  \label{eq:collapsebounds}
\end{align}
where we have expressed all quantities in $d$-dimensional Planck units and neglected order one factors, since they will not be relevant for our discussion. 

Given that the field theory entropy grows like the volume (c.f. eq.  \eqref{eq:standardsethermo}), one could naively try to violate the holographic bounds, \eqref{eq:holographicbounds} by increasing the mass, or equivalently the temperature, of the system. However, we can see from the collapse bounds \eqref{eq:collapsebounds} that before reaching the mass/temperature at which the holographic bound would be violated, the system would gravitationally collapse into a black hole, at least as perceived by an outside observer, which by definition saturates the holographic bound. This can be understood as a gravitational mechanism to \emph{protect} the holographic bounds from being violated \cite{Bousso:2002ju} by field-theoretic configurations. 

Along the previously described process, however, we would observe a sudden increase in the entropy of the system once the box of particles reaches the temperature at which it collapses to a black hole, $T_{\mathrm{coll}}$, since the entropy would jump from $S_{\mathrm{box}}\lesssim\dfrac{L^{d-3}}{T_{\mathrm{coll}}}$ to $S_{\mathrm{BH}}\simeq L^{d-2}$, which is larger since by consistency we have $T\geq L^{-1}$. This discontinuous increase in the entropy makes it difficult to learn anything about the entropy of black holes from that of the tower of particles, apart from giving us a lower bound from the latter. Nevertheless, it can be seen that both bounds scale in the same way precisely if we take the \emph{frozen momentum limit} introduced above eq. \eqref{eq:standardsethermofrozen}, $T\simeq L^{-1}$, such that the entropy of the box of species is given the number of \emph{active} particles in such limit. Therefore, one could hope to get some insight into the microscopic entropy of black holes from that of towers of species if the temperature is increased, keeping $T\simeq L^{-1}$, so that when the system collapses to a black hole its entropy does not jump abruptly. As we will explain, this happens precisely when $T\simeq \Lsp$, and it was the main motivation leading to the formulation of the \emph{Black Hole-Tower Correspondence} in \cite{Herraez:2024kux}, in close analogy to the Black Hole-String Correspondence \cite{Susskind:1993ws,Horowitz:1996nw}, which we review in the following. The latter is deeply related to the original motivation for \emph{Species Thermodynamics} \cite{Cribiori:2023ffn} from the correspondence between the entropy of species and that of \emph{minimal} Black Holes \cite{Basile:2023blg}.

\subsection{The Black Hole-Tower Correspondence}
In this section we first review the Black Hole-String correspondence, as originally presented in \cite{Susskind:1993ws, Horowitz:1996nw}, and then extend it to a general correspondence between towers of species and \emph{minimal black holes}. We focus mainly on weakly coupled string towers and Kaluza-Klein towers, associated to extra dimensions in the decompactification limit, and clear up the concept of \emph{minimal black hole} that probes the species scale in such setups.

\subsection*{The Black Hole-String Correspondence}
\begin{figure}[tb] 
 \centering 
 \includegraphics[width=0.99\textwidth]{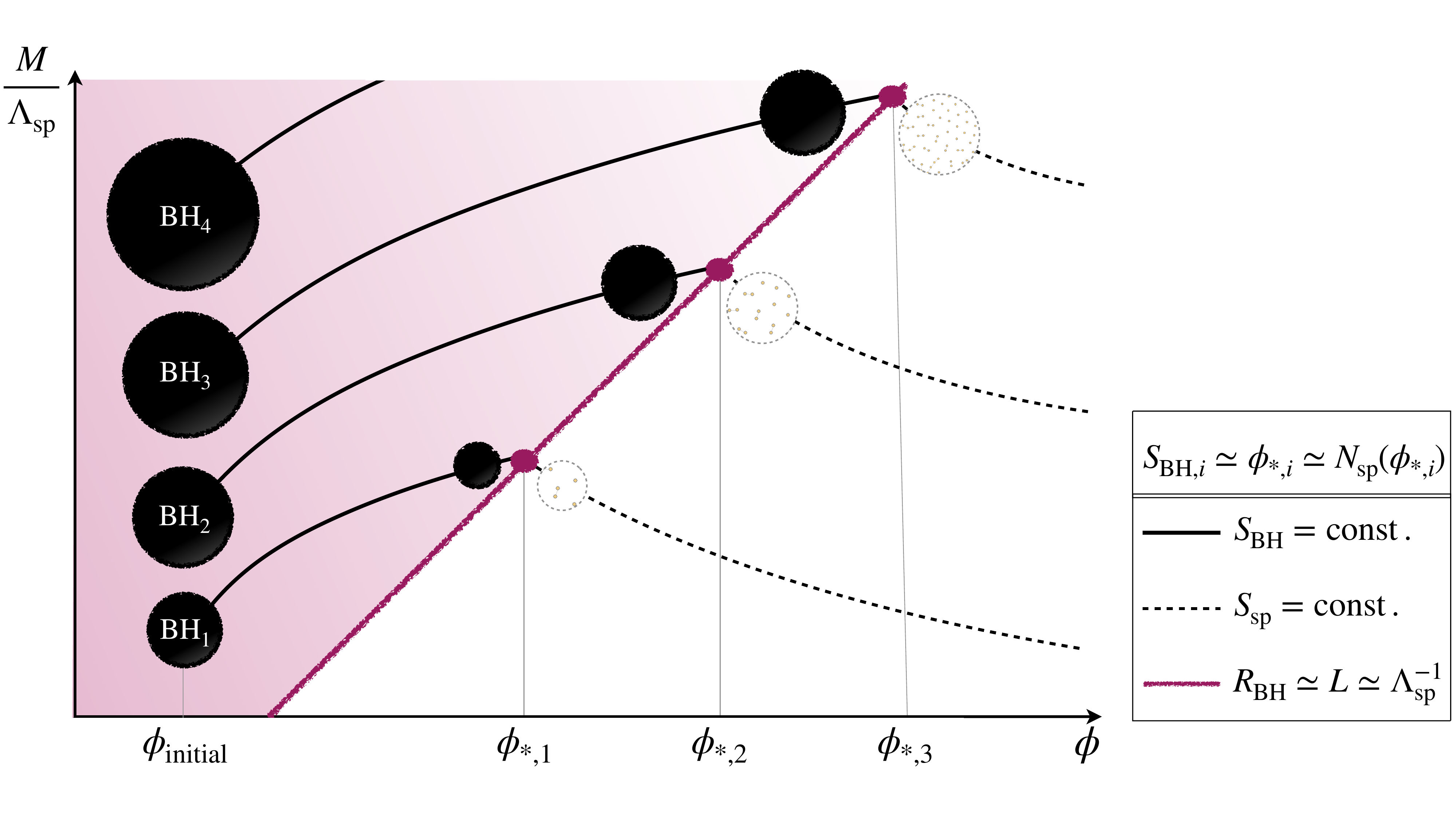} 
 \caption{Constant entropy trajectories in the $M$–$\phi$ plane for several black hole solutions (red lines), with different initial conditions $M$ at a given expectation value $\phi=\phi_0$ of the modulus. At large coupling we depict constant entropy lines for towers of light species (black, dashed lines).The magenta line marks the transition region between the black hole and the tower, defined by $\RBH \sim \ellsp$. Independently of the initial conditions, all configurations can be driven toward this transition region by varying $\phi$.} 
 \label{fig:BHTowercorrespondence}
 \end{figure}

Before presenting the general Black Hole-Tower correspondence, let us briefly review the well-known Black Hole-String correspondence \cite{Susskind:1993ws, Horowitz:1996nw}, which nicely encapsulates the main physical insights and may result more intuitive for the reader familiar with black holes in string theory. To do so, let us consider a toy model string theory in $d$ non-compact dimensions, where the gravitational coupling and the string scale are related through the $d$-dimensional string coupling, $g_{s,d}$, via
\begin{equation}
    \ms^{d-2}=\gsd^2\Mpld^{d-2}\, .
\end{equation}
The mass and entropy of neutral black holes are given in eqs. \eqref{eq:bhentropy}, which we also express here in string units for convenience:
\begin{equation}
\label{eq:MBHSBHstringunits}
    \MBH \simeq \RBH^{d-3}\Mpld^{d-2}\simeq \gsd ^{-2}\RBH^{d-3}\ms^{d-2}\, , \qquad \SBH\simeq\left(\dfrac{\MBH}{\Mpld}\right)^{\frac{d-2}{d-3}}\simeq \gsd^{\frac{2}{d-3}}\left(\dfrac{\MBH}{\ms}\right)^{\frac{d-2}{d-3}} \, .
\end{equation}
If one considers one such black hole and follows it as the string coupling is reduced, making the gravitational interaction weaker and weaker, it is natural to wonder whether at some point the latter could become so weak so as to not able to keep the black hole as a gravitational bound state. In fact, as originally explained in \cite{Susskind:1993ws, Horowitz:1996nw}, it can be seen from eq. \eqref{eq:MBHSBHstringunits} that if we reduce $\gsd$ adiabatically, that is, keeping the entropy of the black hole fixed, its mass in string units increases, but its radius (in the same units) decreases. Interestingly, once the black hole radius is of the order of the string scale, at a value of $\gsd$ that we denote by $g_{d,\ast}$, its mass and entropy are given by
\begin{equation}
\label{eq:MstrSstrcorrespondence}
    M_{\ast}\simeq g_{d,\ast}^{-2}\ms\, , \qquad S_{\ast}\simeq g_{d,\ast}^{-2}\, ,
\end{equation}
which precisely coincides with the entropy of a free string of the same mass. To see this, recall that for a long, free string of length $\Lstr$, its total mass, $M_{\mathrm{str}}$, and entropy, $\Sstr$, are given by (see e.g. \cite{Susskind:1993ws, Horowitz:1996nw} for a derivation of these formulae from a random walk interpretation of the free string)
\begin{equation}
    \label{eq:MstrSstrstringunits}
    M_{\mathrm{str}}\simeq \Lstr\ms^{2}\, \qquad \Sstr\simeq \dfrac{M_{\mathrm{str}}}{\ms}\, .
\end{equation}
so that once one substitutes eq. \eqref{eq:MstrSstrcorrespondence} for the mass it is clear that the entropies coincide. This suggests that once the string scale is reached, above which the black hole solutions are no longer valid, these could become a  string of the same mass, whose entropy can be accounted for in the free limit from its underlying degrees of freedom, hence giving a string theoretic interpretation for the entropy of black holes. Crucially, for any particular initial value of $\gsd\lesssim 1$, all large, neutral black hole can be matched to a corresponding free string as $\gsd$ is decreased adiabatically. The corresponding string is that whose mass at the correspondence point ($\RBH\simeq \ms^{-1}$) matches that of the black hole at the same point, and its entropy $S\simeq g_{d,\ast}^{-2}$ can be thus accounted for by that of the free string, as displayed in Fig. \ref{fig:BHTowercorrespondence} for the general case of the Black Hole-Tower Correspondence explained below. Conversely, for each value of $\gsd\ll 1$, there exists one particular black hole with entropy $\SBH\simeq \gsd^{-2}$. That particular neutral black hole is the one we define as the \emph{minimal black hole} in the context of the species scale---which in this case recovers the string scale---when the tower of species is given by the weakly coupled tower of string oscillators.

\subsection*{The Black Hole-Tower Corespondence}
\label{ss:BHTower}
In order to realize a Black Hole-String correspondence, we must assume the existence of a perturbative string limit---namely a well-defined perturbative regime of quantum gravity characterized by $\ms \ll \Mpld$. As explained in the introduction \ref{s:Intro}, in generic weak-coupling limits of quantum gravity, the species scale can also represent the higher-dimensional Planck scale.

Due to this fact, in the perturbative string regime at the scale $\ms$, stringy effects are known to become important, as the tree-level gravitational contribution becomes comparable to the quantum stringy corrections. This raises the general question of what occurs at energy scales around the species scale.
Hence, inspired by the Black Hole-String correspondence, and by the definition of the species scale \eqref{eq:species_scale}, we can consider a general situation in which the number of species is controlled by a modulus field, $\phi\simeq \Nsp$,\footnote{Notice that strictly speaking the number of species below some scale is a discrete number, but in the weak coupling regimes that we consider here, where this number is sufficiently high, and a thermodynamic analysis is available, we can safely approximate it by a positive real number that varies continuously across the region of interest in field space: $\frac{1}{\Nsp} \ll 1$.} such that the $d$-dimensional Planck scale fulfills
\begin{equation}
    \Lsp^{d-2}=\phi^{-1}\Mpld^{d-2} \, .
\end{equation}
The mass and entropy of a large, neutral black hole can thus be written in terms of the species scale and $\phi$ as
\begin{equation}
\label{eq:MBHSBHspeciesunits}
    \MBH \, \simeq \, \phi\  \RBH^{d-3}\Lsp^{d-2}\, , \qquad \SBH\simeq \dfrac{1}{\phi^{\frac{1}{d-3}}}\left(\dfrac{\MBH}{\Lsp}\right)^{\frac{d-2}{d-3}} \, .
\end{equation}
Thus, by starting with one such large black hole at  some initial value of $\phi$, and then increasing $\phi$ adiabatically, we see that the mass of the black hole in species scale units increases as $\MBH^{d-2}\simeq \phi$ whereas its radius decreases like $R^{(d-3)(d-2)}\simeq \phi^{1-d}$. Precisely when the radius of the black hole becomes of the order of the species scale length, at a value of the modulus that we label $\phi_\ast$, its mass and entropy are given by
\begin{equation}
    M_\ast\simeq \phi_\ast \Lsp\, , \qquad S_\ast\simeq \phi_\ast\, ,
\end{equation}
where we emphasize that $\Nsp=\phi_{\ast}$ at the correspondence point. These are precisely the mass and entropy of the systems of free species considered in section \ref{s:thermo}  precisely when $T\simeq L^{-1}\simeq \Lsp$ (c.f. eq. \eqref{eq:SandEspecies}). A similar picture then arises, in which a large black hole can be connected to a system of free species---whose entropy scales with $\Nsp$---at the correspondence point---when $\RBH\simeq \Lsp^{-1}$---by reducing the $d$-dimensional gravitational coupling by making $\phi$ large. Given a starting value for $\phi$, each large black hole we start with will have a different correspondence point, and thus its entropy can be accounted for by the corresponding system of $\Nsp$ species, with $\Nsp$ given precisely by the number of species. 

Conversely, one can turn this logic around and use the correspondence between black holes and tower of species as a way to determine which kind of towers ought to be allowed in quantum gravity, from requiring that the thermodynamic quantities of the tower ensemble as $T\to\Lsp$ at each point in the space $\phi\ll1$ correspond to those of the \emph{minimal black hole} at the same  $\phi$. This and similar arguments, first presented in \cite{Basile:2023blg} and later rephrased in these terms in \cite{Herraez:2024kux}, have been used to provide a bottom-up argument for the Emergent String Conjecture, and are the main subject of section \ref{s:esc}.

Finally, let us elaborate on the physical meaning of the Black Hole-Tower correspondence for the two types of towers that are known to appear in gravitationally weakly coupled limits, namely the oscillator modes of a fundamental string, or a Kaluza-Klein tower. In the former case, it is easy to recover the Black Hole-String correspondence from our general analysis upon identification of $\Lsp\simeq \ms$ and also $\Nsp=\phi=\gsd^{-2}$, which are known to be the right identifications for emergent string limits \cite{Lee:2019wij}, and thus the \emph{minimal black bole} is simply given by that of size of the order of the string scale in $d$-dimensions.

In the latter case, when applied to Kaluza-Klein towers coming from the decompactification of $p$ dimensions with volume $\mathcal{V}_p$ (measured in higher dimensional Planck units), the right identifications are $\Lsp\simeq M_{\mathrm{Pl,}d+p}$ and $\Nsp=\phi \simeq \mathcal{V}_p$,  which simply come from the usual relation between the higher and lower dimensional Planck scales. In this case, it is illustrative to clarify what we mean by \emph{minimal Black Hole}, namely the one that corresponds to the tower ensemble. By starting with a $d$-dimensional large black hole and following it across constant entropy lines as $\mathcal{V}_p$ is increased,  $\RBH M_{\mathrm{Pl,}d+p}$ is reduced.  Before reaching the species scale, one faces the situation in which $\RBH M_{\mathrm{Pl,}d+p}\simeq \mathcal{V}_p^{1/p}$, namely the horizon size is of the same order as the compactification length scale. From there one, we follow the solution that consists on a higher dimensional black $p$-brane wrapping the volume $\mathcal{V}_p$---which we emphasize is a valid uplift of the original neutral black hole we started with---and continue to shrink the transverse horizon as $\mathcal{V}_p$ becomes larger, until the former approaches the correspondence point, $\RBH M_{\mathrm{Pl,}d+p}\simeq 1$. It is important to remark that this is the only way in which the species scale can be probed by keeping the entropy constant in this setup---for the other possible solution, namely the $(d+p)$-dimensional spherical black hole, the horizon radius is not changed after the point $\RBH M_{\mathrm{Pl,}d+p}\simeq \mathcal{V}_p^{1/p}$ along constant entropy lines, and thus does not probe the species scale.\footnote{We also emphasize that, even though the neutral black $p$-brane solution that wraps the extra dimensions will generically be unstable via e.g. the Gregory-Laflamme instability \cite{Gregory:1993vy}, we can still follow this less entropic solution as a semiclassical saddle of the gravitational action and study its entropy and mass} Thus, the \emph{minimal Black Hole} in the $d$-dimensional EFT for decompactification limits, understood as the one whose horizon size can probe the species scale in the $d$ non-compact dimensions, and whose entropy is given by the number of species at that point, turns out to be a black brane wrapping the compact dimensions. And that configuration is the one whose entropy and mass that can be seen to correspond to those of a system of Kaluza-Klein species in a box of the same transverse size (and also wrapping the extra dimensions), whose entropy is also given by the number of species.

\subsection{Consistency of light towers of states and the Emergent String Conjecture}
\label{s:esc}
In scenarios involving a large number of weakly-coupled fields that interact exclusively through gravity, one might naturally expect that increasing the number of accessible states would inevitably lead to gravitational collapse into a black hole. However, it turns out that the formation of a minimal black hole does not necessarily coincide with the classical expectations from black hole thermodynamics.

In this section, building on the previous results, our primary goal is to review the conditions under which a tower of states exhibits a direct thermodynamic correspondence with a minimal black hole.
The key feature of this correspondence lies in its nature: within the effective field theory description of physics a black hole smaller than the minimal one cannot form. If we aim to generate a larger minimal black hole, the only available mechanism is to increase the number of weakly-coupled species in the universe under consideration. In string theory, this process is typically realized by varying the vacuum expectation value of certain scalar fields known as moduli. 

Let us consider a toy model capturing this equivalence: a weakly-coupled tower of species defined by a characteristic mass scale $\mt$ and a degeneracy spectrum $\mathrm{d}(m)$, both of which may allow for a transition characterized by matching entropy and energy with a minimal black hole.

As one might expect from the loose nature of the constraints, there are infinitely many towers that could realize such a transition. However, the crucial requirement is that the correspondence between the species tower and the minimal black hole must hold for any (large enough) number of species---equivalently, in all asymptotic regions of moduli space. That is, once the nature of the tower and the thermodynamics of the minimal black hole are fixed, we must impose that—despite the differing thermodynamic behavior of the two systems, and hence their varying entropy and energy as a function of the number of species—the transition at the species scale $\Lsp$ must occur universally. This condition is significantly stronger than a mere equivalence between a system of weakly interacting particles and a black hole. Indeed, while large black holes may be understood as coarse-grained ensembles due to their large entropy, minimal black holes impose much tighter constraints on the microscopic nature of the corresponding system.
Within this framework, we ask which towers can be meaningfully interpreted as admitting a transition to a minimal black hole description at the species scale, uniformly in the large $N_{\text{sp}}$ regime. More precisely, we seek to constrain the allowed functional forms of the pairs $(\mathrm{d}_n, \mathrm{f}(n))$ by demanding the thermodynamic matching conditions (see section \ref{s:prelims}):
\begin{align}\label{Constraints1}
E_\text{sp} &= \sum_{n=0}^N \mathrm{d}_n \, m_n = \gamma \, \Lambda_\text{sp}^{3-d} + \mathcal{O}(\Lambda_\text{sp}) \, , \\
S_\text{sp} & = \Lambda_\text{sp}^{2-d} + \mathcal{O}\left(\log \Lambda_\text{sp}^{2-d}\right) \,,
\end{align}
where $\gamma$ identifies an overall $\mathcal{O}(1)$ factor that we have not depicted explicitly in the definition of the species scale \eqref{eq:species_scale}.
However, as discussed above, it is not sufficient to demand that the transition occurs at a single value of the cutoff $\Lsp$ or for a single value of $E_{\text{sp}}$. This stronger condition ensures that the correspondence holds universally across the entire tower and throughout the entirety of asymptotic regions of moduli space, rather than being a feature of a single fine-tuned configuration. Indeed, in some examples, this has been observed to persist in the presence of higher curvature corrections \cite{Cribiori:2023ffn}. Furthermore, even in instances where it seems to be violated at tree level, there is evidence that it may be restored precisely by the inclusion of said higher curvature corrections \cite{Cribiori:2023ffn, Calderon-Infante:2023uhz}.

To begin with, by imposing that the energy of the weakly-coupled tower $E_{\text{sp}}$ matches the mass of the minimal black hole $M_{\text{min, BH}} = \Lsp^{3-d}$ for any number of species $\Nsp$, we can derive the following relation:
\begin{equation}\label{Rec1}
\sum_{n=0}^N \mathrm{d}_n \, \mathrm{f}(n) = \gamma \left(\sum_{n=0}^N \mathrm{d}_n\right) \mathrm{f}(N) \, .
\end{equation}
Next, let us consider moving in moduli space towards the $N_{\text{sp}} \gg 1$ region. This means that the allowed level $N \to N+1$ increases by one, while the degeneracies and masses remain functionally unchanged. However, the energy/mass and entropy will generally change by different amounts as $N_{\text{sp}}$ varies. This implies that we must enforce this equivalence for every value of $N_{\text{sp}}$:
\begin{align}
N'_\text{sp} = \sum_{n=0}^{N+1} \mathrm{d}_n = N_\text{sp} + \mathrm{d}_{N+1}\, , \qquad
\textrm{m}_{N+1} = \Lambda'_\text{sp} = \mt \, \mathrm{f}(N+1).
\end{align}
Applying the same considerations presented above to the rectified tower, we obtain the following recursive equation:
\begin{equation}\label{emergentTowers}
\mathrm{f}(N+1) = \frac{\sum_{n=0}^{N} \mathrm{d}_n}{\sum_{n=0}^{N+1} \mathrm{d}_n - \frac{1}{\gamma} \mathrm{d}_{N+1}} \,\mathrm{f}(N) \quad \Longrightarrow \quad
    \mathrm{f}(N) = \prod_{k=1}^{N-1} \,\frac{\sum_{n=0}^{k} \mathrm{d}_n}{\sum_{n=0}^{k+1} \mathrm{d}_n - \frac{1}{\gamma} \mathrm{d}_{k+1}}.
\end{equation}
This implies that $\mathrm{f}(N+1) > \mathrm{f}(N)$, which is indeed our hypothesis for the existence of a tower, i.e. a non decreasing function $\mathrm{f}(n)$.
According to our argument, the relation in equation \eqref{emergentTowers} defines the only towers that allow a minimal black hole transition at any point within moduli space and indeed the towers parameterized by \eqref{eq:kktower} fulfill this relation. One may wonder what would happen if we chose different ways to reparametrize these towers. Remarkably, even though the individual spectra may change, both the species scale and the functional relationship between the mass spectrum and the degeneracy remain unaffected (see \cite{Basile:2023blg} for a broad class of specific examples illustrating how the constraint operates in practice).
This means that equation \eqref{emergentTowers} effectively encodes different reparametrization of the same underlying tower structure. Namely, the mass spectrum, $\mathrm{d}(m)$, is modified in precisely the exact amount to preserve the form of the species scale $\Lsp$
\begin{align}
     \Lsp \sim  \mt^{\frac{p}{d+p-2}} \,,
\end{align}
where $p$ is a function of the coefficient we use to reparameterize the spectrum.\footnote{As easy example one may consider $\mathrm{d}(n\,;p-1) = \mathrm{d}_0\, n^{p-1}$ which returns $\mathrm{f}(n) = n$. Additionally, if we consider an effective $\mathrm{d}(n) = \mathrm{d}_0$, then \eqref{emergentTowers} returns $\mathrm{f}(n) = n^{\frac{1}{p}}$ which exactly matches with a reparametrization of a Kaluza-Klein tower due to a $p$-cycle compactification. For a more detailed discussion, see \cite{Basile:2023blg}.} This amounts to a transformation of $f(n)$ and $\mathrm{d}(n)$ that yields an effective positive $p$:
\begin{align}
    f(n\,; \gamma, a_i)\,,\, \mathrm{d}(n\,;\gamma, b_i) \longrightarrow{} p = p(\gamma, a_i, b_i) \geq 1\,,
\end{align}
where $a_i$ and $b_i$ collectively denote some extra parameters on which the tower can depend. Crucially, other kinds of towers that cannot be mapped to the one described above do not solve eq. \eqref{emergentTowers} and hence cannot be put in correspondence to a minimal black hole for all $\Nsp\gg 1$.

The criterion discussed in this section is therefore applicable only when the tower of states lies entirely below the UV cutoff, ensuring that the semiclassical expansion of the mass–radius relation remains (at least marginally) valid, i.e. when $\mt \ll \Lsp$\footnote{For such a tower energy fluctuations are not small, so a microcanonical counting is not suitable. A canonical analysis suggests that this corresponds to a Hagedorn transition at $T\simeq \mt$ \cite{Herraez:2024kux}.}.
It is crucial to note that, from the EFT perspective, string states are not light compared to the UV cutoff $\mt = \ms \sim \Lsp$. Consequently, they do not constitute a light tower within the regime of validity of the effective theory. 
Nevertheless, from an EFT viewpoint, there exists the effective limit obtained by taking $p \to \infty$. In this regime, one finds (in Planck units):
\begin{align}
\label{emergentString}
    \Lsp &\sim \mt \,, \\
    E_{\text{sp}} \sim \mt^{3-d} = \Nsp \, \mt\,, \quad
    T_{\text{sp}} &\sim \mt \,, \quad
    S_{\text{sp}} \sim \Nsp = \mt^{2-d}\,.
\end{align}
In this limit, the black hole would undergo a transition to a tower of $\Nsp$ states---all with the energy $\mt$---at a temperature $T \sim \mt$. This behavior is reminiscent and could shed some light into the Hagedorn transition to a free gas of strings with characteristic mass scale $\mt = \ms$ \cite{Atick:1988si, Horowitz:1997jc}.\footnote{Notice that the identification $\Lsp\sim \mt$ not only takes place for the $p\to \infty$ limit, but also for the case $d=2$. This is reminiscent of the observation in \cite{Atick:1988si} that near the Hagedorn temperature the system could behave in a similar way to a 2d CFT.} More explicitly, rewriting this in terms of string quantities yields:
\begin{align}
    \Lsp &\sim \ms \,, \\
    E_{\text{sp}} \sim \frac{\ms}{g_s^2} \,, \quad
    T_{\text{sp}} &\sim \ms \,, \quad
    S_{\text{sp}} \sim \frac{1}{g_s^2} \,,
\end{align}
which precisely matches the Black Hole-String transition  (c.f. eq. \eqref{eq:MstrSstrcorrespondence}) as a limiting case of the Black Hole-Tower transition. We refer to this as the emergent string case.

To recap, whenever these mass spectra satisfy \eqref{emergentTowers}, they are consistent with the Emergent String Conjecture \cite{Lee:2019wij} and reproduce the expected results derived from complementary thermodynamic arguments (cf. eqs. \eqref{eq:SandEspecies}). For a complementary bottom-up argument for the emergent string conjecture see also \cite{Bedroya:2024ubj}.

\section{The laws of Species Thermodynamics}
\label{s:speciest}
\begin{figure}[hb] 
 \centering 
 \includegraphics[width=0.70\textwidth]{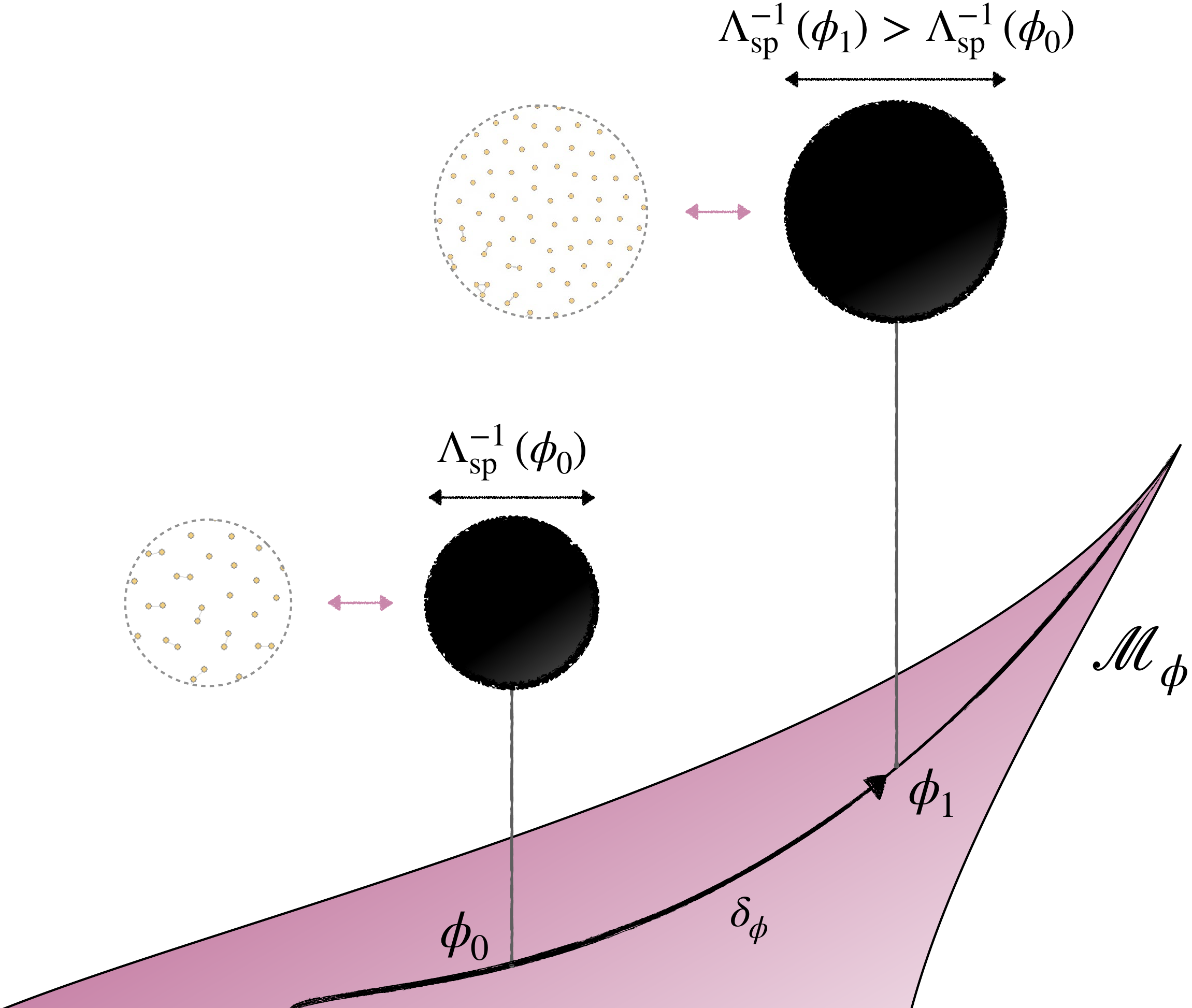} 
 \caption{A pictorial representation of a perturbative corner of a moduli (or field) space $\mathcal{M}_\phi$. For any point $\phi_n$ in $\mathcal{M}_\phi$  we can associate a minimal black hole with radius $\ell_{\text{sp}}(\phi_n) = \Lsp^{-1}(\phi_n) \gg \ell_{\text{Pl},d}$. As one moves towards the boundary of moduli space from $\phi_0$ to $\phi_1$ the species scale must necessarily decrease. Equivalently, we can identify the minimal black holes at each point with radii that increase as one approaches the boundary $\ell_{\text{sp}}(\phi_0) < \ell_{\text{sp}}(\phi_1)$.} 
 \label{fig:species}
 \end{figure} 

As explained in the previous chapters, at any point in the moduli (or field) space $\mathcal{M}_\phi$, we can define a so-called minimal black hole, whose thermodynamic properties depend uniquely on the effective field theory under consideration—that is, on the specific point $\phi$ in field space where the EFT is defined. However, if we want to compare two minimal black holes belonging to two different EFTs, we must establish a way to relate them. Crucially, variations in energy, entropy, and charges cannot be interpreted within a single theory. For example, a minimal black hole cannot increase its energy by absorbing matter from the universe, as it would then no longer be minimal.

Therefore, the only way to modify its thermodynamic properties is to move across different theories—namely, by varying the moduli and traversing the moduli space $\mathcal{M}_\phi$. From this perspective, one may ask whether the laws of black hole thermodynamics also apply to minimal black holes. In this sense, the thermodynamic variation laws should be interpreted as rules that govern the evolution of minimal black holes along paths in $\mathcal{M}_\phi$.

As presented in \cite{Cribiori:2023ffn}, we will reinterpret the laws of black hole thermodynamics relatively to EFTs. We will then introduce the laws of Species Thermodynamics and argue for its validity as a set of principles that must hold in any perturbative region of quantum gravity.

Before proceeding, we first outline some notions that will serve to formalize the upcoming discussion.
Given a point $\phi_0$ in the field space $\mathcal{M}_\phi$, and a curve $\gamma_{\phi_0}$ passing through $\phi_0$, we define the operator $\delta_\phi$ along $\gamma$ as
\begin{equation}
\delta_\phi \, \mathcal{A}(\phi) = \mathcal{A}(\phi + \delta\phi) - \mathcal{A}(\phi)\,,
\end{equation}
where $\mathcal{A}(\phi)$ is a generic function depending both explicitly and implicitly on $\phi$, and $\delta \phi = \left.\frac{d\gamma(\phi(s))}{ds}\right|_\phi \delta s$, with $s$ denoting the curvilinear coordinate along the curve $\gamma$.
In what follows, we restrict our attention to curves $\gamma$ that are geodesics passing through $\phi_0$ and extending to the boundary of the field space $\partial \mathcal{M}_\phi$. The orientation of the curve is defined such that $\delta s > 0$ corresponds to motion toward the boundary.
According to this convention, we define $\phi(s_1) = \phi_1 > \phi(s_0) = \phi_0$ if and only if $s_1 > s_0$, and the distance along the curve $\gamma$ as $\Delta(\phi_0,\phi_1)= \int \left|\frac{d\gamma(\phi(s))}{ds}\right| ds = \int_{\gamma}\sqrt{G_{ij}(\phi)d\phi^i d\phi^j}$.

\subsubsection*{The first law of Species Thermodynamics}

In moduli space $\mathcal{M}_\phi$, given two points arbitrarily close $\phi_0, \, \phi_1 = \phi_0 + \delta \phi $, any two minimal black holes $\mathcal{B}_{\phi_0}, \, \mathcal{B}_{\phi_1}$ are related by\footnote{An extra charge contribution may be added as $\delta_\phi \, E_{\text{sp}} = T_{\text{sp}} \ \delta_{\phi}\, \mathcal{S}_{\text{sp}} \,+  \, \Phi_{\text{sp}}  \, \delta_\phi \mathcal{Q}_{\text{sp}} + \dots$ in presence of charged species \cite{Basile:2024dqq}.}

\begin{equation}
    \delta_\phi \, E_{\text{sp}} = T_{\text{sp}} \ \delta_{\phi}\, \mathcal{S}_{\text{sp}} \, + \dots
\end{equation}

It is important to note that the variation of thermodynamic quantities for this system is not completely free. In the semi-classical regime, within the same EFT, one can in principle modify the mass or entropy of a black hole—for example, by adding mass—yet the first law of black hole thermodynamics enforces a non-trivial relation between these quantities. A minimal black hole, however, corresponds to a distinguished point in black hole phase space: it represents the lightest black hole that may be described whose thermodynamics can be accounted within the effective field theory. Consequently, its thermodynamic properties cannot be altered freely without modifying the underlying number of species—and thus the EFT itself.

The only way to change the thermodynamic data associated with this "transition point" is by moving from a vacuum with moduli $\phi_0$ to another with moduli $\phi_1$. Along such a trajectory, the first law of Species Thermodynamics enforces a correlated variation between the entropy and energy (or mass) of the minimal black hole across two different, but connected, EFTs. Since this black hole also marks the transition to a tower of weakly-coupled species, the constraint must consistently apply to the tower ensemble as well. This was the main idea used in \cite{Basile:2024dqq} to provide a bottom-up motivation of the Emergent String Conjecture \cite{Lee:2019wij}.

\subsubsection*{The second law of Species Thermodynamics}
In moduli space $\mathcal{M}_\phi$, given two  points,   $\phi_0$,  and $ \phi_1 = \phi_0 + \delta \phi$, arbitrarily close to each other, the variation of the species entropy $\delta_\phi \,\mathcal{S}_{\text{sp}}(\phi)$ is non-negative
\begin{equation}
    \forall \phi_0, \phi_1 \in \mathcal{M}_\phi \,, \;  \text{s.t.} \; \min_{\phi \in \partial \mathcal{M}_\phi} \Delta(\phi_1,\phi) > \min_{\phi \in \partial \mathcal{M}_\phi} \Delta(\phi_0,\phi)\,,
\end{equation}
one has
\begin{equation}
    \delta_{\phi} \, \mathcal{S}_{\text{sp}} \geq 0 \Longleftrightarrow
    \delta_{\phi} \, {\Lambda}_{\text{sp}} \leq 0\,.
\end{equation}

Within this perspective, since we are restricted to considering minimal black holes, it is not meaningful to account for external matter falling into them\footnote{The second law of black hole thermodynamics is violated due to quantum effects, namely black holes evaporate by emitting Hawking radiation \cite{Hawking:1974rv}. This can be amended by considering a generalized second law \cite{1974PhRvD...9.3292B}, which extends the mentioned law taking into account the entropy of the matter outside the black hole. This is, at the moment, beyond the scope of such characterization.} Consequently, the (generalized) second law of black hole thermodynamics must be interpreted more abstractly. Once again, any physical change in this system must be accompanied by a change in the underlying EFT. Equivalently, the law must be understood as constrained processes that involves displacements in moduli space.

In this sense, on average, the systems at hand--namely \emph{minimal BHs}, would evolve along the aforementioned constrained trajectories from one to another only in the directions in which the distance to the boundaries of moduli space is reduced or, equivalently, when the species scale is reduced and the number of species increased. In other words, for any displacement between $\phi_0$ and $\phi_1 = \phi_0 + \delta_\phi$, the minimal BH entropy definable within $\mathrm{EFT}_{\phi_1}$ must be greater than or equal to that in $\mathrm{EFT}_{\phi_0}$. Due to the constitutive relations of black hole thermodynamics, this implies that the temperature—and hence the species scale—must decrease along such a process. 

\subsubsection*{The third law of the Species Thermodynamics}
It is impossible, by any physical process, to reduce the species temperature $T_{\text{sp}}$ to zero through a finite sequence of operations $\delta_\phi$. Formally, this can be stated as follows:
\begin{equation}
    \forall\, \mathrm{N} \in \mathbb{N}\,,\; \exists \epsilon > 0 \; \text{s.t.} \; \text{for any non-decreasing sequence} \; \{ \phi_\mathrm{N}\}_{\mathcal{M}_{\phi}}\,,
\end{equation}
one has
\begin{equation}
    \min_{\phi} \Delta(\phi, \phi_\mathrm{N}) > \epsilon \, ,
\end{equation}
The close analogy between the standard laws of thermodynamics and the laws of black hole mechanics is known to break down in the case of the third law, in the sense of the Planck–Nernst formulation as we revisit in the following. Here, we instead consider the so-called weak form of the third law \cite{Wald:1997qp}, which does admit a meaningful analogue in black hole physics—and, as argued in \cite{Bonnefoy:2019nzv}, also holds in the presence of a formal infinite tower of weakly-coupled states.

In this weak form, the third law states that it is not possible to reach the (naïve) state of a black hole at $T = 0$ via any finite sequence of operations. While such a process may be formally defined, it would require an infinite number of steps to be physically realized.

As argued in connection with the other laws, if we apply this reasoning to the minimal black hole within the EFT, the impossibility of connecting a configuration with temperature $T_{\text{sp}}$ to one with $T_{\text{sp}} \to 0^+$ through a finite series of steps implies that such a transformation cannot occur within a finite field displacement $\delta_\phi$ in moduli space. In this sense, the third law of Species Thermodynamics implies that the vanishing species scale defines the infinite-distance limit in moduli space, in alignment with the expectations of the Swampland Distance Conjecture.

\bibliography{refs}
\bibliographystyle{JHEP}

\end{document}